\documentclass[aps,prd,preprintnumbers,superscriptaddress,showpacs,showkeys,floatfix,twocolumn,nofootinbib]{revtex4}

\usepackage{amsmath,amssymb,amsfonts,dcolumn}
\usepackage{graphicx}

\renewcommand{\Im}{\text{Im}\,}

\newcommand{\mpi}{M_{\pi}}

\newcommand{\Fpi}{F_\pi}
\newcommand{\beq}{\begin{equation}}
\newcommand{\eeq}{\end{equation}}
\newcommand{\eps}{\epsilon}

\newcommand{\F}{\mathcal{F}}

\newcommand{\M}{\mathcal{M}}
\newcommand{\Lagr}{\mathcal{L}}
\newcommand{\diff}{\text{d}}

\newcommand{\brho}{\boldsymbol{\rho}}
\newcommand{\bpi}{\boldsymbol{\pi}}
\newcommand{\mr}{M_\rho}
\newcommand{\mrrr}{M_{\rho_3}}
\newcommand{\BR}{\text{BR}}
\providecommand{\keV}{\,\text{keV}}
\providecommand{\MeV}{\,\text{MeV}}
\providecommand{\GeV}{\,\text{GeV}}

\begin{document}

\preprint{INT-PUB-17-041}
\title{Radiative resonance couplings in \boldmath{$\gamma\pi\to\pi\pi$}}

\author{Martin Hoferichter}
\affiliation{Institute for Nuclear Theory, University of Washington, Seattle, WA 98195-1550, USA}

\author{Bastian Kubis}
\affiliation{Helmholtz-Institut f\"ur Strahlen- und Kernphysik (Theorie),
             Universit\"at Bonn,
             D--53115 Bonn, Germany}
\affiliation{Bethe Center for Theoretical Physics,
             Universit\"at Bonn,
             D--53115 Bonn, Germany}
             
\author{Marvin Zanke}
\affiliation{Helmholtz-Institut f\"ur Strahlen- und Kernphysik (Theorie),
             Universit\"at Bonn,
             D--53115 Bonn, Germany}


\begin{abstract}
Studies of the reaction $\gamma\pi\to\pi\pi$, in the context of the ongoing Primakoff program of the COMPASS experiment at CERN,
give access to the radiative couplings of the $\rho(770)$ and $\rho_3(1690)$ resonances. 
We provide a vector-meson-dominance estimate of the respective radiative width of the $\rho_3$, $\Gamma_{\rho_3\to\pi\gamma}=48(18)\keV$, as well as its impact on the $F$-wave in $\gamma\pi\to\pi\pi$.
For the $\rho(770)$, we establish the formalism necessary to extract its radiative coupling directly from the residue of the resonance pole by analytic continuation of the $\gamma\pi\to\pi\pi$ amplitude to the second Riemann sheet, without any reference to the vector-meson-dominance hypothesis.
\end{abstract}



\maketitle

\section{Introduction}

Apart from the two-photon decay of the neutral pion, the process $\gamma\pi\to\pi\pi$ is the simplest manifestation of the Wess--Zumino--Witten anomaly~\cite{Wess:1971yu,Witten:1983tw}.
The leading order in the chiral expansion~\cite{Adler:1971nq,Terentev:1971cso,Aviv:1971hq},
\beq
\label{LET_3pi}
F_{3\pi}=\frac{e N_c}{12\pi^2 \Fpi^3}=9.76(3)\GeV^{-3},
\eeq 
is determined by the number of colors $N_c$, the electric charge $e=\sqrt{4\pi\alpha}$, and the pion decay constant $\Fpi=92.28(9)\MeV$~\cite{Patrignani:2016xqp}. 
Given that early measurements, most prominently $F_{3\pi}=12.9(1.0)\GeV^{-3}$~\cite{Antipov:1986tp}, suggested some tension with the low-energy theorem, 
corrections beyond the leading order~\eqref{LET_3pi} have been worked out~\cite{Bijnens:1989ff,Holstein:1995qj,Hannah:2001ee,Truong:2001en,Ametller:2001yk,Bijnens:2012hf}, with the net result
that higher-order and electromagnetic corrections reduce the value to $F_{3\pi}=10.7(1.2)\GeV^{-3}$.
Together with a similar extraction from $\pi^- e^-\to \pi^- e^- \pi^0$~\cite{Giller:2005uy}, leading to $F_{3\pi}=9.6(1.1)\GeV^{-3}$, the low-energy theorem is now tested at the $10\%$ level, far behind 
the $1.5\%$-level accuracy that has been reached in $\pi^0\to\gamma\gamma$~\cite{Larin:2010kq,Bernstein:2011bx}. 
Meanwhile, a first lattice calculation of $\gamma^*\pi\to\pi\pi$ has been reported in~\cite{Briceno:2015dca,Briceno:2016kkp}.

In contrast to earlier measurements, the Primakoff studies at COMPASS cover not only the threshold region of $\gamma\pi\to\pi\pi$, but extend to much higher center-of-mass energies. As pointed out in~\cite{Hoferichter:2012pm}, this allows one to use the $\rho$ resonance as a lever to vastly increase the statistics of the anomaly extraction, combining constraints from analyticity, unitarity, and crossing symmetry into a two-parameter description of the amplitude whose normalization coincides with $F_{3\pi}$. 
More recently, interest in the $\gamma\pi\to\pi\pi$ reaction has been triggered by its relation to the hadronic-light-by-light contribution to the anomalous magnetic moment of the muon, where it appears as a crucial input quantity for a data-driven determination of the $\pi^0\to\gamma^*\gamma^*$ transition form factor~\cite{Hoferichter:2014vra}, which in turn determines the strength of the pion-pole contribution in a dispersive approach to hadronic light-by-light scattering~\cite{Colangelo:2014dfa,Colangelo:2014pva,Colangelo:2015ama,Colangelo:2017qdm,Colangelo:2017fiz}.

In fact, the kinematic reach of the COMPASS experiment extends up to and including the $\rho_3(1690)$, the first resonance in the $F$-wave. In this paper, we estimate its impact on the $\gamma\pi\to\pi\pi$ cross section based on vector-meson-dominance (VMD) assumptions, which corresponds to an estimate of the radiative width $\Gamma_{\rho_3\to\pi\gamma}$, see Sect.~\ref{sec:VMD}.
For the $\rho(770)$ such a simplified approach is not adequate anymore, precisely due to the amount of statistics available at the $\rho$ peak that should allow one to significantly sharpen the test of the chiral low-energy theorem in the future~\cite{Seyfried}. Instead, the analytic continuation of the $\gamma\pi\to\pi\pi$ amplitudes that underlie this extraction, in combination with the known $\rho$-pole parameters and residues from $\pi\pi$ scattering~\cite{Colangelo:2001df,GarciaMartin:2011jx}, determines the $\rho\pi\gamma$ coupling constant, $g_{\rho\pi\gamma}$, once the free parameters of the representation have been fit to the cross section. The precise prescription for how to extract the radiative coupling of the $\rho$, defined through the residue of the pole in a model-independent way, is spelled out in Sect.~\ref{sec:rho_radiative_coupling}. Combining all currently available information, prior to the direct COMPASS measurement, we predict the line shape of the cross section in Sect.~\ref{sec:line_shape}.
A short summary is provided in Sect.~\ref{sec:summary}.

\section{Vector meson dominance}
\label{sec:VMD}

Throughout, we follow the conventions of~\cite{Hoferichter:2012pm}. The amplitude for the process
\beq
\gamma(q)\pi^-(p_1)\to\pi^-(p_2)\pi^0(p_0)
\eeq
is decomposed according to
\beq
\label{M}
\M_{\gamma\pi\to\pi\pi}(s,t,u)=i\eps_{\mu\nu\alpha\beta}\eps^\mu p_1^\nu p_2^\alpha p_0^\beta \F(s,t,u),
\eeq
in terms of the scalar function $\F(s,t,u)$, the photon polarization vector $\eps^\mu$,
and Mandelstam variables chosen as $s=(q+p_1)^2$, $t=(p_1-p_2)^2$, and $u=(p_1-p_0)^2$, with
$s+t+u=3\mpi^2$, particle masses defined by the charged states, and a relation to the center-of-mass
scattering angle $z=\cos\theta$ according to
\begin{align}
\label{kin_t}
t&=a(s)+b(s) z,\qquad u=a(s)-b(s) z,\notag\\
a(s)&=\frac{3\mpi^2-s}{2},\qquad b(s)=\frac{s-\mpi^2}{2}\sigma_\pi(s),\notag\\
\sigma_\pi(s)&=\sqrt{1-\frac{4\mpi^2}{s}}.
\end{align}
Crossing symmetry implies that the scalar function $\F(s,t,u)$ is fully symmetric in $s,t,u$. In the conventions of~\eqref{M} the cross section becomes
\beq
\sigma(s)=\frac{(s-4\mpi^2)^{3/2}(s-\mpi^2)}{1024\pi\sqrt{s}}\int^1_{-1}\diff z\big(1-z^2\big)|\F(s,t,u)|^2.
\eeq
Later, we also need the partial-wave decomposition~\cite{Jacob:1959at}
\beq
\F(s,t,u)=\sum\limits_{\text{odd }l}f_l(s)P_l'(z),
\eeq
where $P_l'(z)$ denotes the derivative of the Legendre polynomials, and the inversion is given by
\beq
f_l(s)=\frac{1}{2}\int^1_{-1}\diff z\big(P_{l-1}(z)-P_{l+1}(z)\big)\F(s,t,u).
\eeq
Elastic unitarity relates these partial waves to the isospin $I=1$ $\pi\pi$ phase shifts $\delta^1_l(s)$,
\begin{align}
\M_{\pi\pi}^{I=1}(s,t)&=32\pi\sum_{\text{odd }l}(2l+1)t^1_l(s)P_l\bigg(1+\frac{2t}{s-4\mpi^2}\bigg),\notag\\
t^1_l(s)&=\frac{e^{2i\delta^1_l(s)}-1}{2i\sigma_\pi(s)},
\end{align}
by means of
\beq
\label{g3pi_unitarity}
\Im f_l(s)=\sigma_\pi(s)\big(t_l^1(s)\big)^*f_l(s)\theta\big(s-4\mpi^2\big).
\eeq
The fact that the phase of $f_l(s)$ coincides with $\delta^1_l(s)$ is a manifestation of 
Watson's final-state theorem~\cite{Watson:1954uc}. 
Finally, the dominant electromagnetic correction~\cite{Ametller:2001yk} amounts to 
\beq
\F(s,t,u)\to\F(s,t,u)-\frac{2e^2\Fpi^2}{t}F_{3\pi}.
\eeq

\subsection{$\boldsymbol{\rho(770)}$}

The VMD amplitude for $\gamma\pi\to\pi\pi$ can be constructed by combining the $\rho\to\pi\pi$ amplitude from
\beq
\label{rhopipi}
\Lagr_{\rho\pi\pi}=g_{\rho\pi\pi}\eps^{abc}\pi^a\partial^\mu \pi^b\rho_{\mu}^c,
\eeq
with isospin indices $a,\,b,\,c$, together with
\beq
\M_{\rho\pi\gamma}=e g_{\rho\pi\gamma}\eps_{\mu\nu\alpha\beta}\eps^\mu_\rho\eps^\nu_\gamma p_1^\alpha p_2^\beta,
\eeq
where $p_1$ and $p_2$ refer to the momenta of the pion and the photon, and $\eps^\mu_\rho$, $\eps^\nu_\gamma$ to the $\rho$ and $\gamma$ polarization vectors.
The result reads
\beq
\label{f1_VMD}
f_1^\text{VMD}(s)=\frac{2e g_{\rho\pi\gamma}g_{\rho\pi\pi}}{\mr^2-i\mr\Gamma_\rho-s},
\eeq
where the finite width of the $\rho$ has been taken into account by means of a Breit--Wigner propagator.
In Sect.~\ref{sec:rho_radiative_coupling} we will reinterpret both couplings, $g_{\rho\pi\pi}$ and $g_{\rho\pi\gamma}$, as residues of the respective poles, but for the moment we first collect the phenomenological information available when treating the $\rho$ as a narrow resonance. 
In this approximation, the width becomes
\beq
\label{NW_rho}
\Gamma_{\rho\to\pi\pi}=\frac{|g_{\rho\pi\pi}|^2}{48\pi \mr^2}\big(\mr^2-4\mpi^2\big)^{3/2},
\eeq
i.e.\ $|g_{\rho\pi\pi}|\sim 5.95(2)$, with masses and widths as listed in~\cite{Patrignani:2016xqp} (accounting for the different phase space, the results for charged and neutral channels are virtually identical). Similarly, the radiative decay width becomes
\beq
\Gamma_{\rho\to\pi\gamma}=\frac{e^2|g_{\rho\pi\gamma}|^2}{96\pi \mr^3}\big(\mr^2-\mpi^2\big)^3.
\eeq
Within the narrow-width approximation, one could then extract $|g_{\rho\pi\gamma}|$ from the measured cross section for $\gamma\pi\to\pi\pi$ and thereby determine $\Gamma_{\rho\to\pi\gamma}$. At a similar level of accuracy, $SU(3)$ symmetry (see e.g.~\cite{Klingl:1996by}) suggests
$\Gamma_{\rho\to\pi\gamma}= \Gamma_{\omega\to\pi^0\gamma}/9 = 79(2)\keV$, indeed close to 
$\Gamma_{\rho^0\to\pi^0\gamma}= 69(9)\keV$, $\Gamma_{\rho^\pm\to\pi^\pm\gamma}=68(7)\keV$~\cite{Patrignani:2016xqp}.
A model-independent extraction of the radiative coupling of the $\rho$ from $\gamma\pi\to\pi\pi$ will be discussed in Sect.~\ref{sec:rho_radiative_coupling}.

\subsection{$\boldsymbol{\rho_3(1690)}$}
\label{sec:rho_3}

For the generalization to the $\rho_3(1690)$ contribution to $\gamma\pi\to\pi\pi$ and the determination of its radiative width, we follow~\cite{Zanke,Niecknig:2012sj}. 
To this end, we first remark that $G$-parity dictates the photon in this process to have isoscalar
quantum numbers.  In the VMD picture, it therefore couples to the $\rho_3$ and a pion 
predominantly via the $\omega$ meson (assuming the $\phi$ to be negligible due to Okubo--Zweig--Iizuka suppression).
As we aim for a \textit{prediction} of the radiative decay of the $\rho_3$, we in particular need
to assume \textit{strict} VMD without a direct $\rho_3\pi\gamma$ coupling. 
\begin{figure}
\centering
\includegraphics[width=0.5\linewidth]{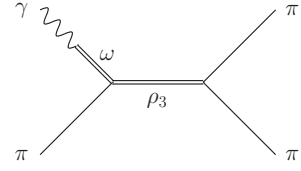}
\caption{VMD mechanism for the $\rho_3$ contribution to $\gamma\pi\to\pi\pi$. \label{fig:g3pi-rho3}}
\end{figure}
Figure~\ref{fig:g3pi-rho3} therefore suggests that we need to determine the coupling constants
$g_{\rho_3\pi\pi}$, $g_{\rho_3\pi\omega}$, as well as $g_{\omega\gamma}$.

Starting from~\cite{Niecknig:2012sj}
\begin{align}
\Lagr_{\rho_3} &= \frac{g_{\rho_3\pi\pi}}{4\Fpi^2} \langle \brho_{\mu\nu\lambda} 
\big[\partial^\mu\bpi,\partial^\nu\partial^\lambda\bpi\big] \rangle \notag\\
&+ \frac{g_{\rho_3\pi\omega}}{2\Fpi} \epsilon^{\lambda\alpha\beta\gamma}
\langle \brho_{\mu\nu\lambda} \partial^\mu \partial_\alpha \bpi \rangle 
\partial^\nu \partial_\beta \omega_\gamma,\notag\\
\Lagr_{\omega\gamma}&= - \frac{e M_\omega^2}{g_{\omega\gamma}}A^\mu\omega_\mu, \label{eq:Lagr-rho3-o-g}
\end{align}
with spin-$3$ fields $\brho_{\mu\nu\lambda} = \rho_{\mu\nu\lambda}^a \tau^a$, pion isotriplet $\bpi = \pi^a\tau^a$, the isoscalar vector field $\omega_\mu$, and the electromagnetic field $A_\mu$, one finds the partial decay widths
\begin{align}
\Gamma_{\rho_3 \to \pi\pi} &= \frac{|g_{\rho_3\pi\pi}|^2}{4480 \pi \Fpi^4\mrrr^2}\big(\mrrr^2-4\mpi^2\big)^{7/2},\notag\\
\Gamma_{\rho_3 \to \pi\omega} &= \frac{|g_{\rho_3\pi\omega}|^2}{13440 \pi \Fpi^2 \mrrr^7}
\lambda\big(\mrrr^2,M_\omega^2,\mpi^2\big)^{7/2},
\end{align}
where $\lambda(x,y,z)=x^2+y^2+z^2-2(x y+x z+y z)$. Together with
$\mrrr = 1688.8(2.1)\MeV$, $\Gamma_{\rho_3} = 161(10)\MeV$, 
$\BR(\rho_3\to\pi\pi) = 23.6(1.3)\%$, and $\BR(\rho_3\to\pi\omega) = 16(6)\%$~\cite{Patrignani:2016xqp}
this fixes the parameters according to
\beq
\label{grho3}
|g_{\rho_3\pi\pi}| = 0.056(2), \qquad
|g_{\rho_3\pi\omega}| = 1.2(2)\GeV^{-2}. 
\eeq
Similarly, one then finds for the radiative width
\beq
\Gamma_{\rho_3 \to \pi\gamma} = \frac{e^2|g_{\rho_3\pi\omega}|^2}{13440 \pi \Fpi^2|g_{\omega\gamma}|^2 \mrrr^7}
\big(\mrrr^2-\mpi^2\big)^{7},
\eeq
and with $|g_{\omega\gamma}|=16.7(2)$ extracted from
\beq
\label{omegaee}
\Gamma_{\omega\to e^+e^-}=\frac{e^4(M_\omega^2-4m_e^2)^{1/2}}{12\pi|g_{\omega\gamma}|^2}\bigg(1+\frac{2m_e^2}{M_\omega^2}\bigg),
\eeq
we obtain the prediction
\beq
\Gamma_{\rho_3 \to \pi\gamma}=48(18)\keV.
\eeq
This result lies slightly higher than the quark-model estimate $\Gamma_{\rho_3 \to \pi\gamma}=21\keV$~\cite{Maeda:2013dka}.
Finally, the resonant contribution to the $\gamma\pi\to\pi\pi$ $F$-wave becomes
\beq
\label{f3_VMD}
f_3^\text{VMD}(s)=\frac{e g_{\rho_3\pi\pi}g_{\rho_3\pi\omega}(s-4\mpi^2)(s-\mpi^2)^2}{60\Fpi^3 g_{\omega\gamma} s (\mrrr^2-i\mrrr \Gamma_{\rho_3}-s)}.
\eeq

\section{Radiative coupling of the $\boldsymbol{\rho(770)}$}
\label{sec:rho_radiative_coupling}

\subsection{$\boldsymbol{\pi\pi}$ scattering}

In a model-independent way, the properties of the $\rho(770)$ are encoded in the pole position and residues of the $S$-matrix on the second Riemann sheet. The prime process to determine 
the parameters is $I=1$ $\pi\pi$ scattering, whose partial-wave amplitude in the vicinity of the pole can be written as
\beq
\label{II_pipi}
t_{1,\text{II}}^1(s)=\frac{g_{\rho\pi\pi}^2(s-4\mpi^2)}{48\pi(s_\rho-s)},\qquad s_\rho=\bigg(M_\rho-i\frac{\Gamma_\rho}{2}\bigg)^2,
\eeq
where the conventions have been chosen in such a way that in the narrow-width limit the coupling $g_{\rho\pi\pi}$ matches onto the Lagrangian definition~\eqref{rhopipi}.
Elastic unitarity for $\pi\pi$ scattering relates the amplitudes on the first and second Riemann sheets according to
\beq
\label{unitarity_pipi}
t_{1,\text{I}}^1(s)-t_{1,\text{II}}^1(s)=-2\sigma^\pi(s)t_{1,\text{I}}^1(s)t_{1,\text{II}}^1(s),
\eeq
where we have introduced~\cite{Moussallam:2011zg}
\beq
\sigma^\pi(s)=\sqrt{\frac{4\mpi^2}{s}-1},\qquad \sigma^\pi(s\pm i\eps)=\mp i \sigma_\pi(s),
\eeq
so that the pole parameters can be determined from the condition that $t_{1,\text{I}}^1(s_\rho)=1/(2\sigma^\pi(s_\rho))=-i/(2\sigma_\pi(s_\rho))$ (since $\Im s_\rho <0$), once a reliable representation of $t_1^1(s)$ on the first sheet is available. 
Such a representation is provided by dispersion relations, in the form of Roy equations~\cite{Roy:1971tc,Ananthanarayan:2000ht,Caprini:2011ky} or variants thereof, the so-called GKPY equations~\cite{GarciaMartin:2011cn}.
The latter produce the pole parameters given in the first line of Table~\ref{tab:rho_pole}, in good agreement with the determination from Roy equations, but with smaller uncertainties. In the following, we use the GKPY parameters from~\cite{GarciaMartin:2011jx} together with the $I=1$ phase shifts from~\cite{GarciaMartin:2011cn}. Within uncertainties, this covers similar determinations listed in the table. 
Note that $g_{\rho\pi\pi}$ is a \textit{complex} coupling, with a phase that is observable (modulo $180^\circ$), although Table~\ref{tab:rho_pole} shows that this phase is rather small~\cite{Jacobo}.

\begin{table}[t]
\renewcommand{\arraystretch}{1.5}
\centering
\begin{tabular}{lrrrr}\toprule
Ref.\ & $M_\rho$ [MeV]  & $\Gamma_\rho$ [MeV] & $|g_{\rho\pi\pi}|$ & $\text{arg}(g_{\rho\pi\pi})$~\cite{Jacobo}\\\colrule
\cite{GarciaMartin:2011jx}, GKPY  & $763.7^{+1.7}_{-1.5}$ & $146.4^{+2.0}_{-2.2}$ & $6.01^{+0.04}_{-0.07}$ & $\big(\!-5.3^{+1.0}_{-0.6}\big)^\circ$\\
\cite{GarciaMartin:2011jx}, Roy  & $761^{+4}_{-3}$ & $143.4^{+3.8}_{-4.6}$ & $5.95^{+0.12}_{-0.08}$ & $\big(\!-5.7^{+1.1}_{-1.4}\big)^\circ$\\
\cite{Colangelo:2001df}, Roy & $762.4(1.8)$ & $145.2(2.8)$ & & \\
\botrule
\end{tabular}
\caption{Pole parameters of the $\rho(770)$ from dispersion relations.
The phase $\text{arg}(g_{\rho\pi\pi})$ in the last column is only determined modulo $180^\circ$.}
\label{tab:rho_pole}
\end{table}

\subsection{Pion form factor}
\label{sec:pion_FF}

The simplest quantity that probes the electromagnetic interactions of the pion is its form factor $F_\pi^V(s)$. Given the wealth of experimental data, it provides an ideal testing ground
to study how well VMD predictions fare when confronted with real data, in this case for $g_{\rho\gamma}$ instead of $g_{\rho\pi\gamma}$.
In analogy to~\eqref{unitarity_pipi}, the elastic unitarity relation,
\beq
\Im F_\pi^V(s)=\sigma_\pi(s)\big(t_l^1(s)\big)^*F_\pi^V(s)\theta\big(s-4\mpi^2\big),
\eeq
defines the analytic continuation of the form factor onto the second sheet
\beq
F_{\pi,\text{I}}^V(s)-F_{\pi,\text{II}}^V(s)=-2\sigma^\pi(s)F_{\pi,\text{I}}^V(s)t_{1,\text{II}}^1(s).
\eeq
In the vicinity of the pole we may write
\beq
F_{\pi,\text{II}}^V(s)=\frac{g_{\rho\pi\pi}}{g_{\rho\gamma}}\frac{s_\rho}{s_\rho-s},
\eeq
where the conventions are chosen in such a way that in the narrow-width and $SU(3)$ limit $g_{\rho\gamma}=g_{\omega\gamma}/3$, cf.~\eqref{eq:Lagr-rho3-o-g}.
Altogether one finds
\beq
\label{FpiV_g}
\frac{1}{g_{\rho\gamma}g_{\rho\pi\pi}}=i\frac{\sigma_\pi^3(s_\rho)}{24\pi}F_{\pi,\text{I}}^V(s_\rho),
\eeq
which allows one to extract $g_{\rho\gamma}$ from the form factor evaluated at $s_\rho$ on the first sheet and the previously determined $g_{\rho\pi\pi}$.
The dispersive formalism for the analytic continuation to $s_\rho$ has been studied in detail in the literature, see~\cite{DeTroconiz:2001rip,Leutwyler:2002hm,Colangelo:2003yw,deTroconiz:2004yzs,Hanhart:2012wi,Ananthanarayan:2013zua,Ananthanarayan:2016mns,Hoferichter:2016duk,Hanhart:2016pcd,Colangelo:2017fiz},
and data abound, mostly motivated by the $\pi\pi$ contribution to hadronic vacuum polarization in the anomalous magnetic moment of the muon.
In this way, the dominant uncertainties actually arise from the error in $g_{\rho\pi\pi}$ as well as the systematics of the fit, e.g.\ whether $\rho$--$\omega$ mixing (as present in the fit to $e^+e^-$ data~\cite{Achasov:2006vp,Akhmetshin:2006bx,Aubert:2009ad,Ambrosino:2010bv,Babusci:2012rp,Ablikim:2015orh}, but not in $\tau\to\pi\pi\nu$~\cite{Fujikawa:2008ma}) is included in the definition of the form factor. 

In the end, the results of the fits fall within the range $|g_{\rho\gamma}|=4.9(1)$, so that, in this case, the VMD expectation, $|g_{\rho\gamma}^\text{VMD}|=|g_{\omega\gamma}|/3=5.6(1)$, agrees with the full result at the $10\%$ level (strict VMD as derived from $\rho\to e^+e^-$ in analogy to~\eqref{omegaee}, without $SU(3)$ assumptions, even produces $|g_{\rho\gamma}^\text{VMD}|=5.0$). The phase comes out around $\text{arg}(g_{\rho\pi\pi}g_{\rho\gamma}) \sim -7^\circ$, so that $g_{\rho\gamma}$ is almost real (with the same sign as the one chosen for $g_{\rho\pi\pi}$).

\subsection{$\boldsymbol{\gamma\pi\to\pi\pi}$}

The derivation for $\gamma\pi\to\pi\pi$ proceeds in close analogy to the pion form factor. From the unitarity relation~\eqref{g3pi_unitarity} we find the analytic continuation
\beq
f_{1,\text{I}}(s)-f_{1,\text{II}}(s)=-2\sigma^\pi(s) f_{1,\text{I}}(s) t_{1,\text{II}}^1(s),
\eeq
and writing
\beq
\label{II_g3pi}
f_{1,\text{II}}(s)=\frac{2e g_{\rho\pi\gamma}g_{\rho\pi\pi}}{s_\rho-s}
\eeq
in the vicinity of the pole (to match onto~\eqref{f1_VMD} in the VMD limit), the analog of~\eqref{FpiV_g} becomes
\beq
\label{g_rhopigamma}
\frac{e g_{\rho\pi\gamma}}{g_{\rho\pi\pi}}=i\frac{s_\rho \sigma^3_{\pi}(s_\rho)}{48\pi}f_{1,\text{I}}(s_\rho).
\eeq

\begin{figure}[t]
 \centering
 \includegraphics[width=\linewidth,clip]{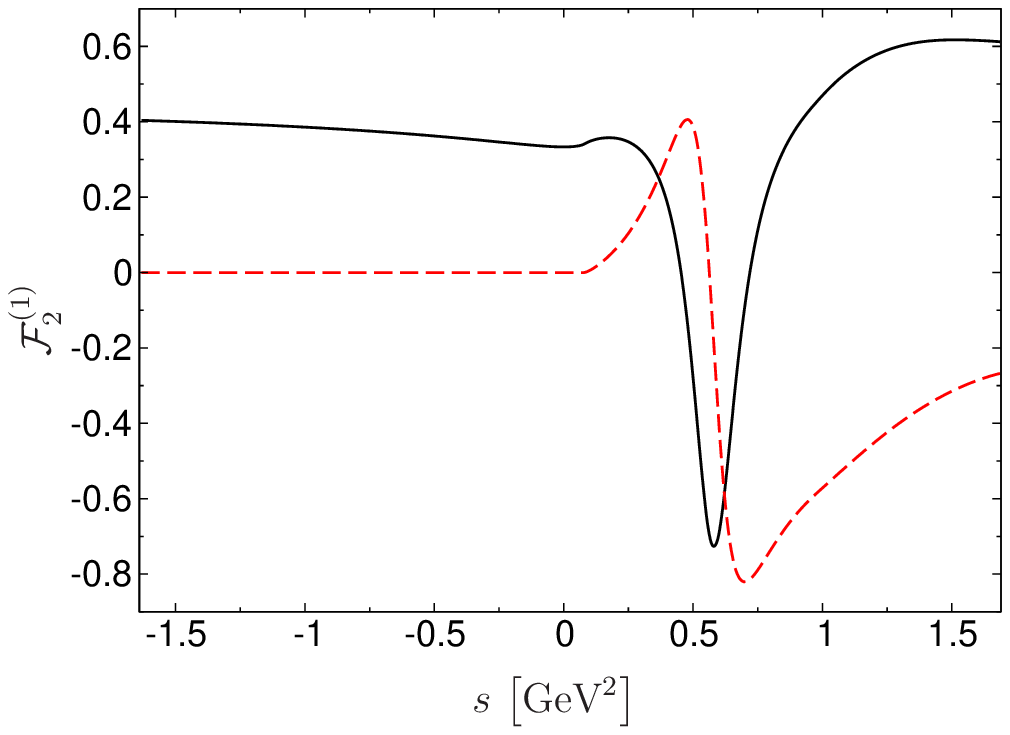}\\
 \includegraphics[width=\linewidth,clip]{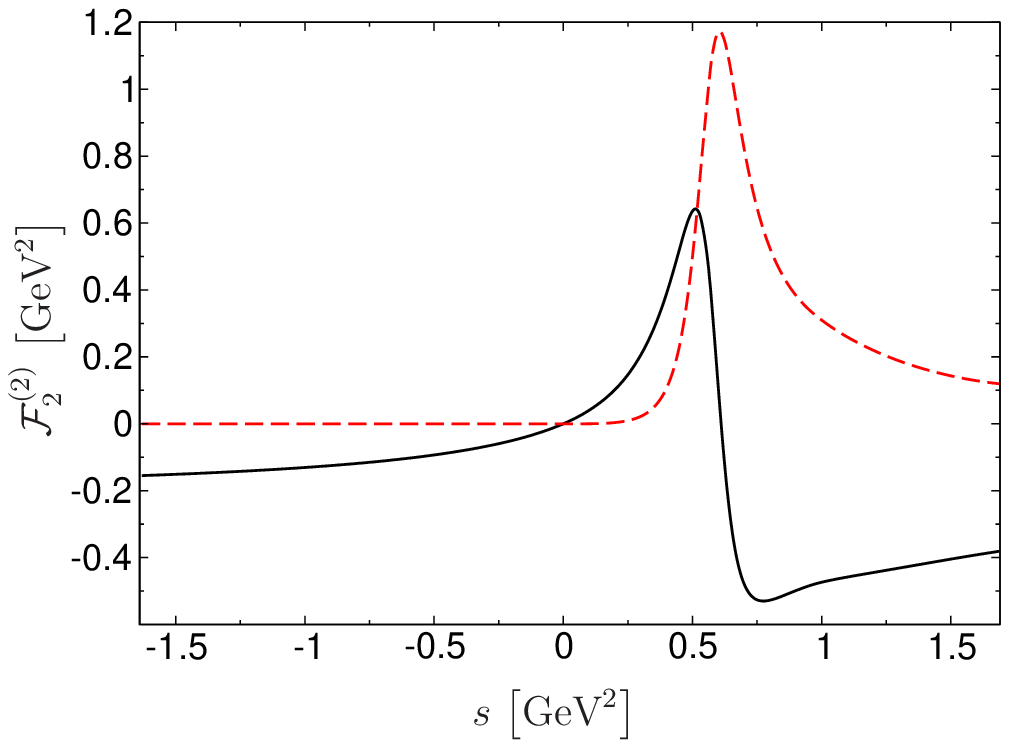}
 \caption{Basis functions $\F_2^{(i)}$ for $\gamma\pi\to\pi\pi$. The black solid (red dashed) lines refer to the real (imaginary) parts.}
 \label{fig:F1F2}
\end{figure}

However, the analytic continuation is less straightforward than for $F_\pi^V(s)$, due to the fact that, in contrast to the form factor, the scattering process $\gamma\pi\to\pi\pi$ produces a left-hand cut, which, in addition, needs to be constructed in such a way that crossing symmetry is maintained. The corresponding formalism has been derived in~\cite{Hoferichter:2012pm}. Starting from the decomposition
\beq
\label{F_symm}
\F(s,t,u)=\F(s)+\F(t)+\F(u),
\eeq
which holds if imaginary parts from partial waves with $l\geq 3$ are neglected, it was shown that the solution of the dispersion relation for $\F(s)$ can be represented in the form
\begin{align}
\label{F_twice}
\F(s)&=C_2^{(1)}\F_2^{(1)}(s)+C_2^{(2)}\F_2^{(2)}(s)\notag\\
&=\frac{1}{3}\big(C_2^{(1)}+C_2^{(2)}s\big)+\frac{1}{\pi}\int_{4\mpi^2}^\infty\frac{\diff s'}{s'^2}\frac{s^2}{s'-s}\notag\\
&\qquad\times
\big(C_2^{(1)}\Im\F_2^{(1)}(s')+C_2^{(2)}\Im\F_2^{(2)}(s')\big),
\end{align}
where $C_2^{(i)}$ refer to the subtraction constants in the twice-subtracted dispersion relation. These are the free parameters of the fit. In contrast, the basis functions $\F_2^{(i)}(s)$ can be calculated once and for all, for a given input of the $\pi\pi$ phase shift $\delta_1^1(s)$ (the results for the phase shift from~\cite{GarciaMartin:2011cn} are depicted in Fig.~\ref{fig:F1F2}).
The partial wave $f_1(s)$ follows from
\begin{align}
 f_1(s)&=\frac{3}{4}\int_{-1}^1\diff z\big(1-z^2\big)\big(\F(s)+\F(t)+\F(u)\big)\notag\\
 &=C_2^{(1)}+C_2^{(2)}\mpi^2+\frac{1}{\pi}\int_{4\mpi^2}^\infty\diff s'\, K(s,s')\notag\\
 &\qquad\times\big(C_2^{(1)}\Im\F_2^{(1)}(s')+C_2^{(2)}\Im\F_2^{(2)}(s')\big),
\end{align}
with integration kernel
\begin{align}
K(s,s')&=\frac{s^2}{s'^2(s'-s)}+\frac{3}{b(s)}\Big\{\big(1-x_s^2\big)Q_0(x_s)+x_s\Big\}\notag\\
&-\frac{2}{s'}+\frac{s-3\mpi^2}{s'^2},\qquad x_s=\frac{s'-a(s)}{b(s)},
\end{align}
and the lowest Legendre function of the second kind
\begin{align}
Q_0(z)&=\frac{1}{2}\int^1_{-1}\frac{\diff x}{z-x},\notag\\
Q_0(z\pm i\eps)&=\frac{1}{2}\log\bigg|\frac{1+z}{1-z}\bigg|\mp i\frac{\pi}{2}\theta\big(1-z^2\big).
\end{align}
For the GKPY $\rho$ parameters from~\cite{GarciaMartin:2011jx} we obtain
\begin{align}
f_{1,\text{I}}(s_\rho)&=C_2^{(1)}\big(0.588(5)+0.193(7)i\big)\notag\\
&\qquad-C_2^{(2)}\big(0.071(7)+0.570(5)i\big)\GeV^2,
\end{align}
where the uncertainties reflect the propagated errors on the pole parameters only (when experiment reaches few-percent accuracy, also the uncertainties in the $\pi\pi$ phase shift will have to be included).
Once the $C_2^{(i)}$ are fit to cross-section data, this relation determines $f_{1,\text{I}}(s_\rho)$, and thus, by means of~\eqref{g_rhopigamma}, the radiative coupling of the $\rho(770)$
(including its phase).
The current knowledge of these couplings, see~\eqref{C_central} below, indicates that, similar to $g_{\rho \gamma}$, $g_{\rho \pi \gamma}$ is almost real with the same sign as $g_{\rho \pi \pi}$. 

\section{Line shape of $\boldsymbol{\gamma\pi\to\pi\pi}$}
\label{sec:line_shape}

Currently available information on the radiative coupling of the $\rho(770)$~\cite{Patrignani:2016xqp} largely derives from the high-momentum Primakoff experiments~\cite{Jensen:1982nf,Huston:1986wi,Capraro:1987rp}, while no experimental result is available for the $\rho_3(1690)$ at all. Thanks to its high-statistics data, COMPASS has the unique opportunity 
to determine these couplings either for the first time or with unprecedented accuracy; compare their results for the radiative widths of the $a_2(1320)$ and the $\pi_2(1670)$ as extracted from the similar Primakoff reaction $\gamma\pi\to3\pi$~\cite{Adolph:2014mup}. For the $\rho(770)$ such a measurement is intimately related to the determination of the chiral anomaly, and, building upon~\cite{Hoferichter:2012pm}, the previous section establishes the formalism to extract both simultaneously in a consistent, model-independent way. 

In this section, we reverse the argument and collect the currently available information to predict the line shape to be expected in the $\gamma\pi\to\pi\pi$ cross section. First of all, the 
combination $C_2^{(1)}+C_2^{(2)}\mpi^2$ is related to the chiral anomaly, but only up to an additional quark-mass renormalization
\beq
C_2^{(1)}+C_2^{(2)}\mpi^2=\bar F_{3\pi}\equiv F_{3\pi}\big(1+3\mpi^2\bar C\big),
\eeq
estimated from resonance saturation to $3\mpi^2\bar C=6.6\%$~\cite{Bijnens:1989ff}. We use the corresponding central value, but, given that we wish to extract $\bar F_{3\pi}$ from the data, assign a $10\%$ uncertainty, 
$\bar F_{3\pi}=10.4(1.0)\GeV^{-3}$, to reflect the level of accuracy that previous measurements have established.
The second combination of coupling constants corresponds to the radiative coupling of the $\rho(770)$, for which we take the $SU(3)$ VMD result $|g_{\rho\pi\gamma}|=0.79(8)\GeV^{-1}$, but, in view of the results for the pion form factor in Sect.~\ref{sec:pion_FF}, attach a $10\%$ uncertainty as well. These constraints translate into
\beq
\label{C_central}
C_2^{(1)}=9.9(1.0)\GeV^{-3},\qquad C_2^{(2)}=24.1(2.5)\GeV^{-5},
\eeq
where the uncertainty in $C_2^{(1)}$ and $C_2^{(2)}$ is entirely dominated by $\bar F_{3\pi}$ and $|g_{\rho\pi\gamma}|$, respectively.
For the $\rho_3(1690)$ we use the parameters as given in Sect.~\ref{sec:rho_3}.

The twice-subtracted dispersion relation~\eqref{F_twice} is perfectly suited to extract the coefficients $C_2^{(i)}$ from cross-section data up to and including the $\rho$ resonance, but displays a pathological high-energy behavior. To obtain a description that remains valid in the whole region below $2\GeV$, we implement a version of the basis functions $\F_2^{(i)}$ with a relatively low cut-off parameter $\Lambda=1.3\GeV$ that leaves the low-energy physics virtually unaffected, but allows us to introduce a high-energy completion of the resulting partial wave $f_1(s)\sim 1/s$, in agreement with general arguments based on the Froissart bound~\cite{Froissart:1961ux}.
In addition to the $P$-wave, the symmetrized version~\eqref{F_symm} produces non-vanishing contributions to $f_l(s)$ for all (odd) $l$ in the partial-wave projection. However, we checked that the corresponding $F$- and higher partial waves can be ignored, and similarly effects from excited $\rho$ states, $\rho'$ and $\rho''$, are likely negligible~\cite{Zanke,Schneider:2012ez} (assuming that these resonances couple with a comparable relative strength as in the pion form factor~\cite{Fujikawa:2008ma}).
While inelastic corrections included directly in the dispersive description by means of the inelasticity parameter are typically small~\cite{Niecknig:2012sj}, 
such excited $\rho$ states provide an indicator for the size of the dominant inelasticities from $4\pi$ intermediate states; see also~\cite{Hanhart:2012wi}.\footnote{We wish to emphasize that our dispersive representation of the $P$-wave is very reliable mostly below $1\GeV$, and the model for the $F$-wave around the $\rho_3$ resonance. Despite the indications for comparably smaller $\rho'$, $\rho''$ contributions, we do not claim to make a high-precision prediction of the line shape between $1$ and $1.5\GeV$.} 
Finally, the narrow-resonance approximation for the $\rho_3$ is strictly meaningful only at the resonance mass, while the additional momentum dependence in~\eqref{f3_VMD} distorts the resonance shape. To obtain a more realistic line shape we follow~\cite{VonHippel:1972fg,Adolph:2015tqa} and introduce centrifugal-barrier factors, which amounts to the replacement
\beq
 f_3^\text{VMD}(s)\to f_3^\text{VMD}(s)\frac{B_3(q_\text{f}(s)R)B_3(q_\text{i}(s)R)}{B_3(q_\text{f}(\mrrr^2)R)B_3(q_\text{i}(\mrrr^2)R)},
\eeq
with $B_3(x)=15/\sqrt{225+ 45x^2 + 6x^4+x^6}$,
initial- and final-state momenta
\beq
q_\text{i}(s)=\frac{s-\mpi^2}{2\sqrt{s}}, \qquad q_\text{f}(s)=\sqrt{\frac{s}{4}-\mpi^2},
\eeq
and a scale $R\sim 1\,\text{fm}$. 
\begin{figure}[t]
 \centering
 \includegraphics[width=\linewidth,clip]{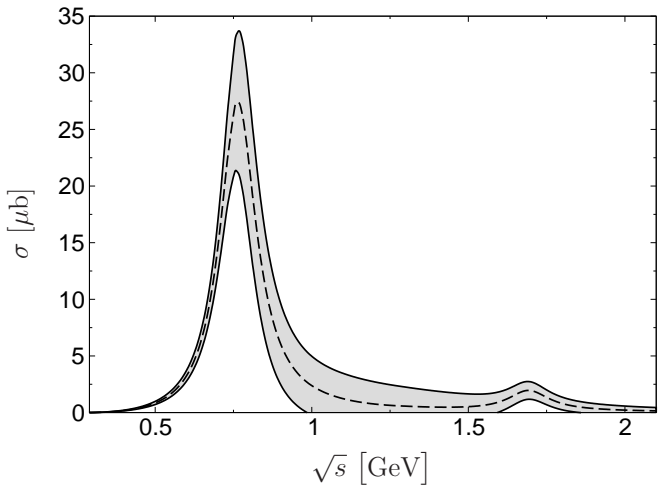}
 \caption{Total cross section for $\gamma\pi\to\pi\pi$. The dashed line refers to our central solution, using~\eqref{grho3} and~\eqref{C_central}.}
 \label{fig:cross_section}
\end{figure} 
The resulting cross section is shown in Fig.~\ref{fig:cross_section}.  Note in particular that the peak of the $F$-wave cross section,
\begin{align}
\sigma_F\big(\mrrr^2\big) &= \frac{3\big(\mrrr^2-4\mpi^2\big)^{3/2}\big(\mrrr^2-\mpi^2\big)}{896\pi\mrrr} \big|f_3\big(\mrrr^2\big)\big|^2 \notag\\
&= \frac{56\pi\mrrr^2}{\big(\mrrr^2-\mpi^2\big)^2} \frac{\Gamma_{\rho_3\to\pi\pi}\Gamma_{\rho_3\to\pi\gamma}}{\Gamma_{\rho_3}^2}\notag\\
&= 1.7(6)\,\mu\text{b},
\end{align}
amounts to roughly 6\% of the dominant $\rho(770)$ peak.
While currently the uncertainties are large, an improved measurement of the energy dependence would immediately translate to better constraints on the underlying QCD parameters, most notably the chiral anomaly $F_{3\pi}$, but also, as we have shown in this paper, the radiative couplings of the $\rho(770)$ and $\rho_3(1690)$ resonances.

\section{Summary}
\label{sec:summary}

Extending the dispersive formalism for $\gamma\pi\to\pi\pi$ developed in~\cite{Hoferichter:2012pm}, we have worked out the analytic continuation necessary to extract the radiative coupling of the $\rho(770)$, as defined by the residue at its resonance pole. Throughout, we have indicated the correspondence to the parameters that would occur within a narrow-resonance description, and collected the current phenomenological information. Combined with a VMD estimate for the $\rho_3(1690)$, we have obtained a prediction for the cross section of $\gamma\pi\to\pi\pi$ up to $2\GeV$, with uncertainties dominated by the current knowledge of the underlying parameters: the chiral anomaly $F_{3\pi}$ and the radiative couplings of the $\rho(770)$ and $\rho_3(1690)$ resonances. 

This prediction can be considered a benchmark for the ongoing Primakoff program at COMPASS. Measuring the cross section with reduced uncertainties compared to Fig.~\ref{fig:cross_section} would allow one to test the narrow-width estimate of the $\rho_3\to\pi\gamma$ decay rate, $\Gamma_{\rho_3\to\pi\gamma}=48(18)\keV$, and to improve the determination of the chiral anomaly and the $\rho\to\pi\gamma$ coupling, both without relying on model assumptions while still profiting from the full statistics of the $\rho$ resonance. Such improved experimental information on $\gamma\pi\to\pi\pi$ is particularly timely given its relation to hadronic light-by-light scattering in the anomalous magnetic moment of the muon as well as recent lattice calculations. We look forward to the results of the ongoing analysis of the $\pi^-\pi^0$ channel at COMPASS that will provide an important step in this direction.

\begin{acknowledgments}
We would like to thank Jan Friedrich, Boris Grube, Misha Mikhasenko, Stephan Paul, Jacobo Ruiz de Elvira, and Julian Seyfried for helpful discussions. 
Financial support by  the DOE (Grant No.\ DE-FG02-00ER41132) and the DFG (CRC 110 ``Symmetries and the Emergence
of Structure in QCD'') is gratefully acknowledged.
\end{acknowledgments}

\end{document}